# INITIAL EXPERIMENTAL RESULTS OF A MACHINE LEARNING-BASED TEMPERATURE CONTROL SYSTEM FOR AN RF GUN

A.L. Edelen[#], S.G. Biedron, S.V. Milton, Colorado State University, Fort Collins, CO, USA
B.E. Chase, D.J. Crawford, N. Eddy, D. Edstrom Jr., E.R. Harms, J. Ruan, J.K. Santucci, Fermi National Accelerator Laboratory*, Batavia, IL, USA
P. Stabile, ADAM, Geneva, Switzerland

*Abstract*

Colorado State University (CSU) and Fermi National Accelerator Laboratory (Fermilab) have been developing a control system to regulate the resonant frequency of an RF electron gun. As part of this effort, we present initial test results for a benchmark temperature controller that combines a machine learning-based model and a predictive control algorithm. This is part of an on-going effort to develop adaptive, machine learning-based tools specifically to address control challenges found in particle accelerator systems.

## INTRODUCTION

The electron gun at the advanced superconducting test accelerator [1-3] is a 1½ cell normal-conducting copper RF photoinjector operating at 1.3 GHz. It is water-cooled and shows a 23-kHz shift in resonant frequency per °C change in cavity temperature. Thus, establishing satisfactory control of the water temperature at the cavity entrance is the first step toward ensuring the gun is kept at the proper resonant frequency. Existing requirements state that this water temperature should be regulated to within ±0.02°C [2]. This regulation loop can then be nested within another control algorithm that determines what the water temperature needs to be in order to either a) directly minimize the detuning or b) achieve an operator-specified cavity temperature set point. As an intermediate result, this discussion considers the latter case. This also facilitates comparison with the existing controller.

### Water System Overview

A simplified schematic of the water system is given in Fig. 1. A detailed description is given in Ref. [4]. The two controllable variables are 1) the flow control valve setting and 2) the heater power setting. For this particular system there are several control challenges:

- Due to water transport and thermal time constants, long time delays exist in the system responses (~10s from the valve to T02, ~30s from T02 to TIN, ~20s from TIN to TCAV, and ~60s from TOUT to T06).
- Without compensation, any change in the temperature of the water exiting the gun (either due to an increase in waste heat from the gun or a change in the temperature of the water entering the gun) will circulate back into the mixing chamber and have a secondary impact on the cavity temperature.
- There are fluctuations in the low conductivity water (LCW) supply temperature. While it is nominally regulated to within ±0.5°C, larger spikes can occur.

Due to the TCAV sensor location and the cavity geometry, the temperature recorded will be higher than the cavity wall temperature under RF power. Thus, for resonance control using operator-specified TCAV set points, it is important to note that the set point required to maintain the proper resonant frequency will increase with increasing average RF power.

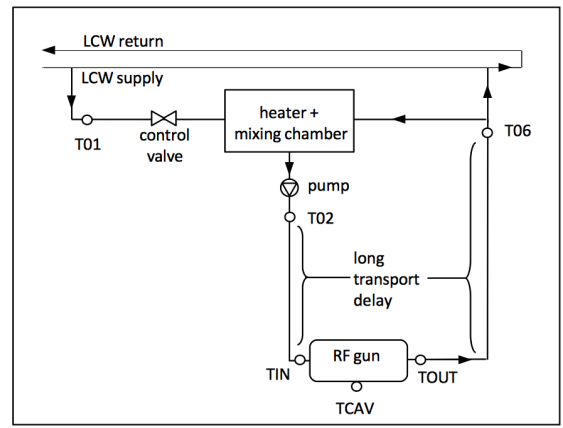

Figure 1: Layout of the water system and relevant instrumentation. T01, T02, TIN, TCAV, TOUT, and T06 are temperature sensors.

### Existing Controller

Presently, the cavity temperature is being regulated by a feedforward/PI controller developed at Fermilab that adjusts the valve setting such that a TCAV set point is reached. An older version of the controller is described in [4], and a recent response to a 1-°C step change in the set point is shown in Fig. 2. This is under no RF power.

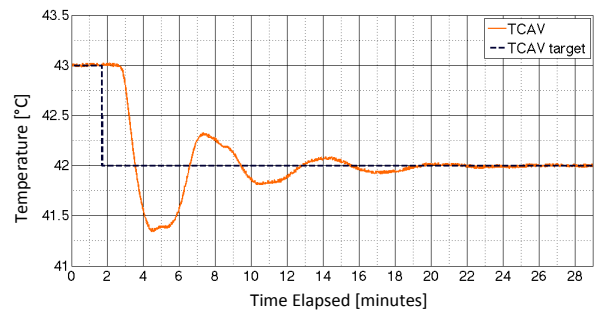

Figure 2: A 1-°C step change under the existing PI controller. The oscillations are due to water recirculation.

The oscillations and long settling time are due to recirculation of the water (and its associated temperature changes) through the system. In the instance shown, it takes ~23 minutes to reach steady state. Note that this is without significant disturbances to the supply temperature. While this is acceptable for largely stable operations at current average power levels, a significant improvement in settling time, overshoot, and disturbance rejection could be gained by adopting alternative control techniques. This has implications for machine up-time and management of reflected power.

## DESIGN CONSIDERATIONS

Because of the long time constants, the effect of the water returning from the gun, and the presence of two controllable variables, a model-based predictive control (MPC) scheme was chosen. In MPC, a system model and an optimization algorithm are used to determine an optimal sequence of future controller actions such that the target output is reached within some future time horizon, subject to the satisfaction of any defined constraints. Figure 3 shows the basic concept of MPC.

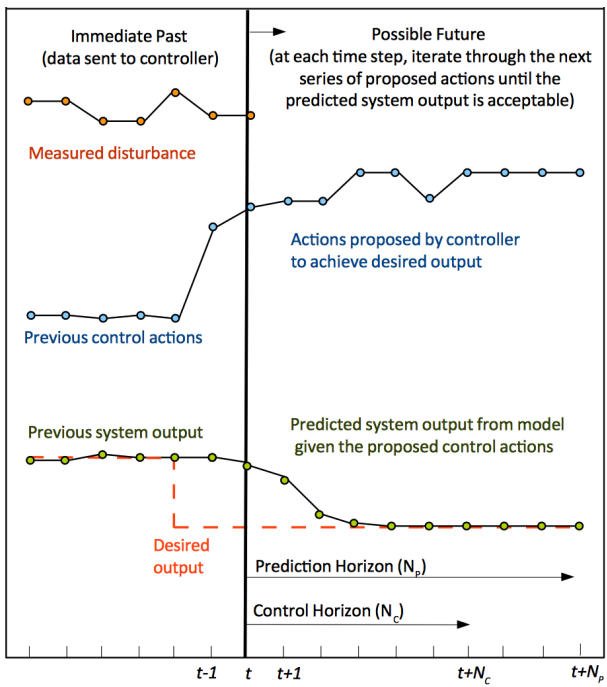

Figure 3: Basic concept of model predictive control.

Given the system layout, it makes sense to have the main MPC unit regulate the temperature immediately after the mixing chamber (i.e. at T02). By monitoring T01 and using this as a model input, adjustments can be made to compensate for fluctuations in the LCW supply temperature. By also monitoring the temperature of the water leaving the gun, the controller can compensate for any changes before they reach the mixing chamber. Monitoring TOUT provides plenty of time for actuation of the heater to take effect (whereas monitoring T06 does not). Finally, if a series of future set points is known in advance, the controller can act anticipatively to reduce the effect of dead time in the system.

While elements of the system can be modeled analytically, a model that is developed using measured data helps to ensure the plant-model mismatch is kept relatively small. To this end, a neural network model of the heater/mixing chamber subsystem was designed. This model takes in 20s of relevant (i.e. with dead time removed) previous values of T01, TOUT, valve position, and heater power to predict the next value of T02.

## BENCHMARK MPC

To guide future design work, a performance benchmark for temperature regulation using a simple quadratic programming formulation of MPC was desired. This helps to define what kind of performance we can expect without having to grapple just yet with the tradeoff between solution quality and computation time that comes with using more sophisticated MPC designs.

To ensure good solutions for the optimization problem could be reached within the 2-s control interval, the model was linearized. Over test data, the root mean squared error of the simplified model is 0.073°C, whereas for the original model it is 0.008°C. No communication or actuation delays were taken into account during model training.

Finally, a rudimentary neural network model that takes a user-specified TCAV set point and yields the appropriate T02 set point was also created. Figure 4 shows a conceptual diagram of the benchmark MPC.

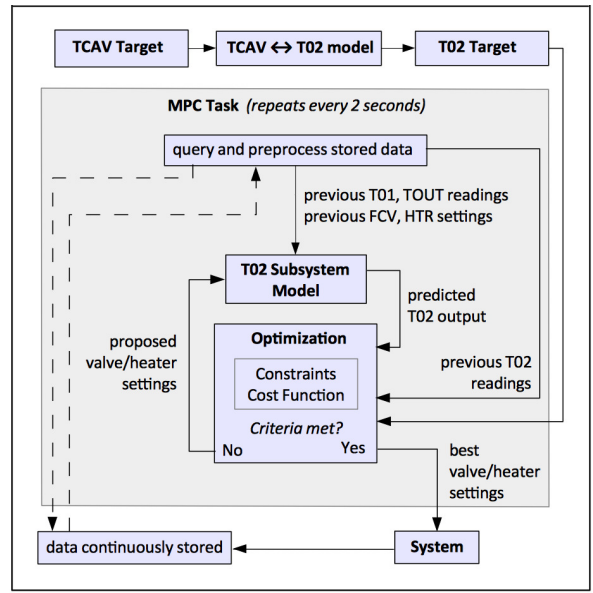

Figure 4: Conceptual diagram for the benchmark MPC.

Candidate MPCs with various prediction/control horizons, constraints, and specific cost functions were examined via simulation and testing. Some parameters of the benchmark MPC are shown in Table 1.

Note that T02 was upgraded between the model design and the controller design, resulting in a temperature offset and improved noise characteristics. An approximate static offset was incorporated into the data preprocessing for T02, but because the model was trained on the noisier data it is not quite as sensitive to small changes in the input parameters as it could be with some re-training.

Table 1: Benchmark MPC Parameters

| Parameter | Value | Units |
|---|---|---|
| Valve max rate | 10 | [% open] |
| Valve upper limit | 70 | [% open] |
| Valve lower limit | 2 | [% open] |
| Heater max rate | 4 | [kW] |
| Heater upper limit | 8.9 | [kW] |
| Heater lower limit | 1 | [kW] |
| Prediction horizon | 100 | [s] |
| Control horizon | 20 | [s] |
| Control interval | 2 | [s] |

*Benchmark MPC Performance*

A 1-°C step in cavity temperature is shown in Fig. 5. Note that the scales are smaller than those shown in Fig. 2 (1.5°C vs. 2.5°C and 10min vs. 30min), and once again there is no RF power going to the gun. After the step command for the cavity is issued, the MPC brings T02 to within ±0.02°C of its respective set point in about 3 minutes. Correspondingly, TCAV is brought to within ±0.02°C of its set point in about 5 minutes.

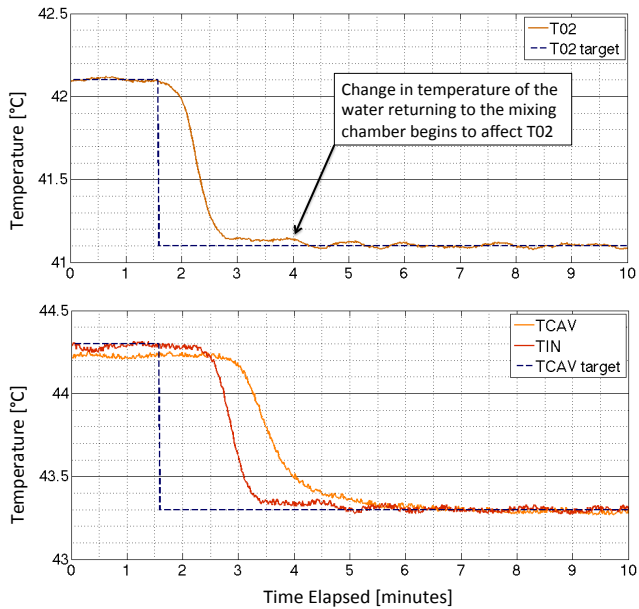

Figure 5: 1-°C change in TCAV under the benchmark MPC. Note that the scales are smaller than those of Fig. 2.

The small oscillations in T02 that start at the 4-minute mark are the result of imperfect timing in the compensative actions for the recirculating water.

The small steady state offset in TCAV after the step is likely due to modeling error between the TCAV and T02 set points. This portion of the controller contained no feedback to account for such steady state errors. In addition to modeling errors, TIN and TCAV had not yet completely reached a steady state prior to the step.

Figure 6 shows the measured valve and heater actions. We see an initial adjustment (the valve opens and the heater power decreases), followed by an adjustment to compensate for the lower temperature of the water exiting the gun. Note that the requested actions (not shown) are slightly offset in timing and are a bit smoother than the readbacks.

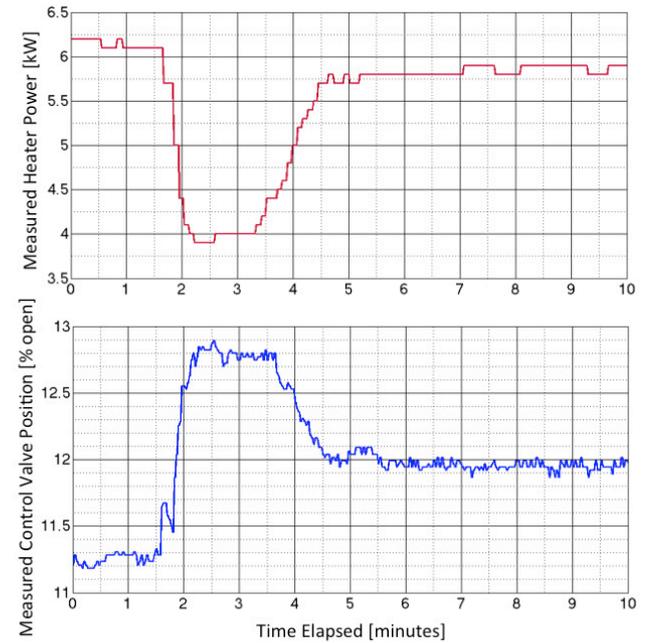

Figure 6: Measured flow control valve and heater actions.

This is a significant improvement over the performance achieved with more conventional control techniques. Additional work is still needed before the controller can be used reliably at different RF power levels. The T02 model performs well under powered conditions, and thus in principle the MPC should be able to compensate for temperature changes in the water exiting the gun associated with RF power adjustments. However, the component that converts the TCAV set point to a T02 set point needs to be more carefully designed before this is implemented for regular use.

## CONCLUSIONS

The benchmark MPC was able to adjust the valve and heater settings during a 1-°C step such that the TCAV set point was reached to within ±0.02°C in about 5 minutes. In the process, the control actions compensated for temperature changes in the recirculating water.

We are now confident that building a more sophisticated MPC is a sensible way forward for the water temperature control. We also plan to extend the controller to regulate the resonant frequency directly using the water temperature and measurements of the cavity RF signals.

## REFERENCES

[1] Fermilab-TM-2568, 2013.
[2] M. Church (Ed.), "Design of the Advanced Superconducting Test Accelerator", Fermilab beams-doc-4212, 2012.
[3] P. Piot, et al., TUPME041, Proc. IPAC14, (2014).
[4] P. Stabile, et al., TUPSM13, Proc. PAC13, (2013).